\documentclass[aps,twocolumn,nofootinbib]{revtex4}

\usepackage{graphicx}
\usepackage{amssymb, amsmath,mathtools}
\usepackage{bbm}
\usepackage{enumerate}

\allowdisplaybreaks

\usepackage{xcolor}

\usepackage{bm}   
\usepackage{dsfont}   

\usepackage{hyperref} 
\hypersetup{
colorlinks = true,   
urlcolor = blue,    
linkcolor = magenta,   
citecolor = blue  
}
\usepackage{float}
\usepackage{color}
\begin{document}

\title{Phases of three-dimensional $d$-wave superconductors with two attractive interactions}

\author{Majid Kheirkhah$^1$ and Igor F. Herbut$^{1,2}$}
\affiliation{$^1$ Department of Physics, Simon Fraser University, Burnaby, British Columbia V5A 1S6, Canada \\  $^2$ Institute for Solid Sate Physics, University of Tokyo, Kashiwa 277-8581, Japan}

\begin{abstract}
Three-dimensional $d$-wave ($j=2$) superconducting state may result from Cooper pairing in channels with different angular momenta and spin, such as $(l=0, s=2)$ or $(l=2, s=0)$. We consider spin-3/2 Luttinger fermions in the limit of small spin-orbit coupling parameter and with weak attractive interactions in both  of these channels, and demonstrate that the stable $d$-wave superconducting phase of the system below critical temperature depends critically on the ratio of two interactions. When a weak inversion-breaking term in the kinetic energy is included, all three of the real, ferromagnetic, and cyclic phases can be realized within a narrow range around unit ratio. The result shows how models of multicomponent superconductors can be surprisingly sensitive to precise values of some of its parameters and display radically different superconducting ground states with their variation.
\end{abstract}

\maketitle

\section{Introduction}

Cooper pairing of effective-spin-3/2 particles which can appear in trapped alkali fermions \cite{ho} and spin-orbit-coupled three-dimensional materials \cite{ butch, bay, kim, boettcher1, meinert, brydon, savary, boettcher2, venderbos, roy, mandal, link, Ipsita} has attracted plenty of attention over the years. Prominent among many possibilities is the $d$-wave superconducting state with the total angular momentum of $j=2$. It can arise in many ways; for orbital momenta $l<3$, for example, the  state can be realized by pairing with (1) the orbital momentum $l=0$ and total spin $s=2$, (2) the orbital momentum $l=2$ and spin $s=0$ or $s=2$, and (3) $l=1$ and $s=1$ or $s=3$. Any such $d$-wave state is described by five complex order parameters which transform as an irreducible second-rank tensor under rotational group SO(3). The lowest-energy superconducting phase is well known to depend on two quartic coefficients in the Ginzburg-Landau (GL) free energy \cite{mermin, sauls, ho, kawaguchi}. The phase diagram features in principle three superconducting phases: two, the cyclic and the ferromagnetic phases, break time-reversal symmetry, while the real superconducting phase preserves it. Particularly intriguing is the cyclic phase, which, despite breaking the time-reversal symmetry exhibits zero average magnetization. In contrast, the ferromagnetic superconducting phase has maximal magnetization. The precise form of the real phase is ultimately determined by the higher-order (sextic) terms in the GL free energy \cite{boettcher2}.

An essential new feature of the higher-spin pairing is that, as mentioned above and in general, superconducting states in the same representation of the relevant symmetry group can arise by pairing in more than one channel. If the quasiparticle interaction happens to be attractive in two (or more) channels, then the resulting  macroscopic superconducting state could be a linear superposition of different pairing states. Naturally, when this occurs the coefficients in the superposition will depend on the relative strengths of different pairing interactions. This raises the following question: Does the ultimate stable superconducting phase depend on the precise superposition that is dictated by the pairing interactions? Phrased more technically: Could the quartic and/or higher-order terms in the GL theory that ultimately select the stable superconducting phase depend on the particular superposition dictated by the quadratic term in a nontrivial way that would allow different stable phases for different superpositions?

In this work we consider this issue in probably the simplest setting of two attractive interactions in the orbital, $(l=2, s=0)$,  and spin, $(l=0, s=2)$, $d$-wave channels for spin-3/2 Luttinger fermions, at finite chemical potential, and in the limit of small spin-orbit (band-inverting) coupling parameter. Additional motivation for choosing this example is that in the two limiting cases of attractions in just one of the two channels it is known that one of the quartic GL coefficients [$q_3$, see Eq. (\ref{eq_quartic}) below] is exactly zero \cite{ho, boettcher2, majid1}, which causes an accidental degeneracy between the cyclic and the ferromagnetic phases at this truncation of the GL theory. We therefore first set to check if this coefficient becomes finite when the two $j=2$ states form a superposition. To our surprise, we find that this coefficient remains zero even for a general superposition, and, as far as we can tell,  the accidental degeneracy between the ferromagnetic and cyclic phases persists. We, nevertheless, find that the other quartic coefficient ($q_2$) {\it changes sign} as the superposition angle is varied and therefore allows the time-reversal symmetry to be both preserved and violated by the superconducting state. Finally, we perturb the Luttinger particle propagator with a weak rotationally invariant term linear in momentum, which explicitly breaks the inversion symmetry. This is found to produce a finite coefficient $q_3$ and, furthermore, to do so in precisely such a way that allows {\it all three} of the above phases to be traversed as the superposition angle is swept over its full range. In particular, the cyclic state --- a unique state in the $j=2$ Hilbert space that breaks time-reversal symmetry but has zero average magnetization --- is identified as a possible ground state of the Luttinger Hamiltonian. The question posed above is therefore answered in the affirmative. One can expect similar phenomenon occurring in other systems that involve multicomponent superconducting order parameter, although counter examples, which we discuss further in the concluding section, also exist \cite{link}.

A complementary issue to the one raised here was studied in Ref. \cite{ho}, where it was shown that the $j=2$ superconducting ground state in the systems with separate rotational symmetries in orbital and spin spaces depends on the value of the spin of the particles that are being paired in the $l=0$, $s=2$ channel. For spin-3/2 and spin-5/2 particles the stable $d$-wave state is real, whereas for spin higher or equal to 7/2 it is ferromagnetic. The cyclic phase, however, in this spin-orbit-decoupled system does not occur.

The paper is organized as follows. In sec. II we define the system under study: the quadratically dispersing four-component Luttinger fermions, attracting each other in two different $j=2$ channels. In sec. III we discuss the GL free energy, mixing of the two order parameters, and the general phase diagram. We present our main results for the three coefficients of the quartic terms of the GL free energy  as functions of the mixing angle in sec. IV. The discussion of the results is given in sec. V. Details of the calculation are provided in two Appendixes.

\section{Attractive Luttinger fermions}

We define the system of Luttinger fermions with attractive interactions in two $j=2$ channels by the action $\mathcal{S} = \int _0 ^{1/T} d\tau \mathcal{L}$, with the Lagrangian \cite{boettcher1, boettcher2, mandal}
\begin{widetext}
\begin{align*}
\mathcal{L} = \sum_ {\bm{p}}  \psi^{\dagger}(\bm{p}, \tau)
\big[\partial_{\tau} + \mathcal{H}_0(\bm{p}) \big]
\psi (\bm{p}, \tau)
-
\sum_{\bm{p}, \bm{q}} \sum_{s = 0,2}
V_s
\big[
 \psi^{\dagger} (-\bm{p}, \tau)  \gamma_{45} \mathcal{A}_a^{(s)}(\bm{p}) \psi^{\ast} (\bm{p}, \tau)
 \big]
 \big[
\psi^{\rm T}(-\bm{q}, \tau)  \mathcal{A}_a^{(s)}  (\bm{q}) \gamma_{45} \psi  (\bm{q}, \tau)
\big],
\end{align*}
\end{widetext}
where the summation in the last term is over the index $a = 1,2,\dots, 5$, $\bm{p} = (p_x,p_y,p_z)$ is the momentum,  $\psi= (c_{\frac{3}{2}}, ~c_{\frac{1}{2}},~c_{-\frac{1}{2}},~c_{-\frac{3}{2}})^{\rm T}$ is the four-component (spin-3/2) Grassmann field, $\tau$ represents imaginary time,  $\mathcal{A}_a^{(0)} (\bm{p}) = \frac{d_a(\bm{p})}{p^2} \mathds{1}_{4\times 4}$, and $\mathcal{A}_a^{(2)} (\bm{p}) = \gamma_a$ and independent of momentum. $\gamma_a$ are $4\times 4$ Hermitian matrices, and $ \mathds{1}$ is the $4\times 4$ unit matrix. The positive couplings $V_s$ for $s=0,2$ represent attractive interactions in the ($l=2$, $s=0$) and ($l=0$, $s=2$) channels, respectively. The interaction is assumed to be rotationally invariant separately in the orbital and in the spin spaces. As will become clear later, this assumption can be relaxed and the spin-orbit coupling allowed in the pairing term, with essentially no change in the conclusions.

The five Hermitian gamma matrices $\gamma_a$ as usual obey the Clifford algebra $\{\gamma_a, \gamma_b \} = 2\delta_{ab}$, and will here be chosen to be $\gamma_1 = \sigma_1 \otimes  \mathds{1}_{2 \times 2}$, $\gamma_2 = \sigma_3 \otimes \sigma_3$,
$\gamma_3 = \sigma_3 \otimes \sigma_1$, $\gamma_4 = \sigma_3 \otimes \sigma_2$, $\gamma_5 = \sigma_2 \otimes \mathds{1}_{2\times2} $, where $\sigma_i$ ($i=1,2,3$) are the usual Pauli matrices. The unitary part of the time-reversal symmetry operator is then $\gamma_{45} = i \gamma_4 \gamma_5$ \cite{boettcher1}. The real $l=2$ spherical harmonics $d_a(\bm{p})$ are defined as
\begin{align}
d_1(\bm{p}) &= \frac{\sqrt{3}}{2}(p^2_x - p^2_y),
\hspace{5mm}
d_2(\bm{p}) = \frac{1}{2}(3p^2_z - p^2),
\nonumber
\\
d_3(\bm{p}) &= \sqrt{3} p_x p_z,
\hspace{1.5mm}
d_4(\bm{p}) = \sqrt{3} p_y p_z,
\hspace{1.5mm}
d_5(\bm{p}) = \sqrt{3} p_x p_y.
\nonumber
\end{align}
The rotationally symmetric and inversion-symmetric single-particle Luttinger Hamiltonian \cite{luttinger, abrikosov, janssen} reads as
\begin{equation}
\mathcal{H}_0(\bm{p}) = (p^2- \mu)\mathds{1}_{4 \times 4} + \lambda \sum_{a=1}^5  d_a(\bm{p})\gamma_a,
\end{equation}
where $\mu >0$ is the chemical potential. The parameter $\lambda$ measures the strength of spin-orbit coupling. Here we  will be interested in the small spin-orbit coupling limit when $\lambda \ll 1$, so that both bands disperse in the same direction, arbitrarily chosen to be upward. We neglect the anisotropy which would reduce the full rotational symmetry down to cubic symmetry \cite{boettcheranisotropy} to keep the calculation that follows, which is already quite involved, not overly complicated.

The complex order parameters
\begin{align}
\Delta^{(s)}_a =  \sum_{\bm{p}}
\Big \langle
 \psi^{\rm T}(-\bm{p}, \tau) \gamma_{45} \mathcal{A}_a^{(s)} (\bm{p})  \psi (\bm{p},\tau)
\Big \rangle,
\end{align}
for $s \in \{0,2\}$ characterize five-component $d$-wave superconductivity with Cooper pairs with orbital angular momentum and spin $(l,s) = (2,0)$ and $(0,2)$, both having the total angular momentum $j = l+s = 2$.

The single-particle propagator that corresponds to the Luttinger Hamiltonian is the standard
\begin{align}
G_0(Q) &=
\big[ i \omega_n \mathds{1}_{4 \times 4}  - \mathcal{H}_0(\bm{p}) \big]^{-1}
\nonumber
\\&=
\frac{(i \omega_n  - p^2 + \mu) \mathds{1}_{4 \times 4} + \lambda d_a(\bm{p}) \gamma_a}{(i \omega_n  - p^2 + \mu)^2 - \lambda^2 p^4},
\label{prop_1}
\end{align}
where we introduced $Q =(\bm{p},\omega_n)$ and $\omega_n = (2n+1)\pi T$ is the fermionic Matsubara frequency at temperature $T$ for $n\in \mathbb{Z}$.

\section{Ginzburg-Landau free energy}

The second-order term in the GL free energy $F= F_2 + F_4$ for a uniform $d$-wave order parameter is given by
\begin{equation}
F_2(\Delta_a)  =
\hspace{-2mm}
\sum_{a=1}^5
\hspace{-1mm}
\begin{pmatrix}
\Delta_a^{*(0)} &  \Delta_a^{*(2)}
\end{pmatrix}
\mathcal{F}_{ab}(x)
\begin{pmatrix}
\Delta_b ^{(0) } \\  \\ \Delta_b ^{(2) }
\end{pmatrix},
\end{equation}
where the real, symmetric, $2 \times 2$ temperature-dependent matrix $\mathcal{F}_{ab}(x)$ reads as (see Appendix \ref{app_A})
\begin{align}
\mathcal{F}_{ab}(x) =
\begin{pmatrix}
V_0 ^{-1} + \mathcal{N}  x
&&&
-x y
\\
-x y
&&&
V_ 2 ^{-1} + \mathcal{N}  x
\end{pmatrix}
 \delta_{ab} .
\label{F_matrix}
\end{align}
Here, $x = -\frac{2}{5}\log(\frac{\Omega}{T})$, with $\Omega$ as the energy cutoff,
$\mathcal{N} = \mathcal{N}_{+}(\mu,\lambda) + \mathcal{N}_{-}(\mu,\lambda)$ is the total density of states at the Fermi level, where
\begin{align}
\mathcal{N}_{\pm}(\mu,\lambda) =
\frac{\sqrt{\mu}}{2 \pi^2
(1 \pm \lambda)^{3/2}},
\end{align}
for $0 \leqslant \lambda < 1$, and
\begin{align}
y = \mathcal{N}_{-}(\mu,\lambda) - \mathcal{N}_{+}(\mu,\lambda) \simeq \frac{3 \sqrt{\mu } }{2 \pi ^2}
\lambda + \mathcal{O}(\lambda^3),
\end{align}
is a small positive parameter that is responsible for the mixing between two superconducting states with $(l,s) = (2,0)$ and $(0,2)$. Had we allowed the spin-orbit coupling in the interaction term as well, there would be an additional constant off-diagonal term which would amount to a redefinition of the parameter $y$.

As the temperature decreases, one of the eigenvalues of matrix $\mathcal{F}_{ab}(x)$ first becomes zero, corresponding to the critical temperature $T_c$. In other words, $T_c$ is the highest temperature for which ${\rm det}(\mathcal{F}_{ab}(x)) = 0$. Solving this quadratic equation yields
\begin{align}
x_0 = \frac{\sqrt{\mathcal{N}^2(V_0 - V_2)^2 + 4V_0 V_2 y^2} -\mathcal{N}(V_0 + V_2)}{2V_0 V_2 (\mathcal{N}^2 - y^2)},
\label{eq_x0}
\end{align}
and $T_c = \Omega \exp (5x_0 /2)$.  Right below $T_c$, the superconducting order parameter becomes a  linear combination
\begin{align}
\Delta_a = \Delta^{(0)}_a \cos \alpha
+ \Delta^{(2)}_a \sin \alpha,
\label{linear_comb}
\end{align}
where the ``superposition angle" $ 0 \leqslant \alpha \leqslant  \pi/2$ satisfies the eigenvector equation
\begin{align}
\mathcal{F}_{ab}(x_0)
\begin{pmatrix}
\cos \alpha \\  \sin \alpha
\end{pmatrix}
=0.
\end{align}
Equivalently,
\begin{align}
\cot \alpha = \dfrac{x_0 y}{V_0 ^{-1} + \mathcal{N} x_0}.
\label{eq_cot}
\end{align}

The next, fourth-order (quartic) term in the GL free energy for the $d$-wave order parameter may conveniently be written as \cite{boettcher2, mermin}
\begin{align}
F_4(\Delta_a) &= q_1 [{\rm tr}(\phi \phi^{\dagger})]^2
+ q_2 {\rm tr}(\phi \phi)
{\rm tr}(\phi^{\dagger} \phi^{\dagger})
\nonumber
\\&\quad
+ q_3{\rm tr}(\phi \phi^{\dagger} \phi \phi^{\dagger}),
\label{eq_quartic}
\end{align}
where the five order parameters $\Delta_a$ are used to form the  traceless symmetric tensor $\phi = \sum_{a=1}^5 \Delta_a M_a$, with $M_a$ ($a=1,2,\dots,5$) as real matrices that provide a basis in the space of symmetric, traceless, $3 \times 3$ matrices:
\begin{align}
M_1 &=
\begin{pmatrix}
1 & 0 & 0 \\
0 & -1 & 0 \\
0 & 0 &0
\end{pmatrix},
~
M_2 = \frac{1}{\sqrt{3}}
\begin{pmatrix}
-1 & 0 & 0 \\
0 & -1 & 0 \\
0 & 0 & 2
\end{pmatrix} ,
\nonumber
\\
M_3 &=
\begin{pmatrix}
0 & 0 & 1 \\
0 & 0 & 0 \\
1 & 0 &0
\end{pmatrix},
~
M_4 =
\begin{pmatrix}
0 & 0 & 0 \\
0 & 0 & 1 \\
0 & 1 &0
\end{pmatrix} ,
\nonumber
~
M_5 =
\begin{pmatrix}
0 & 1 & 0 \\
1 & 0 & 0 \\
0 & 0 & 0
\end{pmatrix}.
\end{align}

We assume hereafter that $q_1 >0$, as will be also found shortly in the actual calculation. Depending on the signs and relative magnitudes of the ratios $q_{2}/q_{1}$ and $q_{3}/q_{1}$, the lowest free energy is then found in either the real, cyclic, or ferromagnetic state \cite{mermin, kawaguchi, link2}. When $q_3 =0$, the sign of $q_2$ alone determines  whether the superconducting ground state violates or preserves time-reversal symmetry. For $q_2>0$, the order parameter with $|{\rm tr} (\phi \phi)|=2 |\Delta_a \Delta_a| = 0$ is clearly favorable, as it leads to the largest condensation energy. This state breaks time-reversal symmetry maximally, and it turns out \cite{mermin} can be either the cyclic state with the minimal value of ${\rm tr} (\phi \phi^\dagger \phi \phi^\dagger)= (1/3) ({\rm tr} \phi \phi^\dagger)^2 $ when $q_3 >0$,  or the ferromagnetic state with the maximal value of ${\rm tr} (\phi \phi^\dagger \phi \phi^\dagger)= ({\rm tr} \phi \phi^\dagger)^2 $ when $q_3 <0$. For $q_2<0$ and $q_3 =0$, on the other hand, the maximal value of $|\Delta_a \Delta_a|$ becomes favorable, which implies a real (modulo overall phase) $\Delta_a$ \cite{remark}. This superconducting state would preserve the time-reversal symmetry, and for any such real state  ${\rm tr} (\phi \phi^\dagger\phi \phi^\dagger) = (1/2) ({\rm tr} (\phi \phi^\dagger ))^2$ \cite{link2}. The degeneracy between the real states is lifted by higher-order terms in the GL free energy. As the value of $|q_3|$ increases, in the region with $q_2 < 0$  there is eventually a transition from the real state into either the cyclic or the ferromagnetic state.

Next, one integrates out the Luttinger fermions in presence of superconducting order to compute one-loop BCS expressions for the GL coefficients \cite{ho, boettcher2}. The details of this procedure \cite{zwerger, herbut} are standard and given in the Appendix. In order to obtain a finite coefficient $q_3$, we perturb the Luttinger Hamiltonian by adding the term $\delta \bm{p}\cdot\bm{J}$ to the numerator of the propagator [given by Eq.~(\ref{prop_1})] with small parameter $\delta$, where $\bm{J} = (J_x, J_y, J_z)$ are the usual spin-$\frac{3}{2}$ generators of the SO(3) (see Appendix \ref{app_B}). The term  $\sim \bm{p}\cdot \bm{J}$ is still rotationally invariant but, unlike the rest of the Luttinger Hamiltonian, is odd under inversion, since the  momentum $\bm{p}$ is a vector and the angular momentum $\bm{J}$ is a pseudovector. We may therefore consider this modification of the Luttinger propagator as the computationally simplest way to introduce explicit breaking of the inversion symmetry in the problem. It transpires, however, that any such term that violates inversion symmetry and introduces odd powers of angular momentum $j=3/2$ generators into the Luttinger propagator would yield a finite coefficient $q_3$ and have qualitatively similar effects. We return to this discussion in the concluding section.

\section{Results}

As summarized in Appendix \ref{app_B}, at low temperatures and for small values of both spin-orbit coupling $\lambda$ and inversion symmetry-breaking parameter $\delta$, we find the GL quartic coefficients as functions of the superposition angle $\alpha$ to the leading order to be
\begin{align}
q_i = \frac{\mathcal{N}_0 }{T_c ^2}
\Big[
a w_i(\alpha) +
\Big( b u_i(\alpha) +  c v_i(\alpha) \Big)
\frac{\mu \delta^2}{T_c ^2}
\Big],
\label{eq_qi}
\end{align}
where $\mathcal{N}_0 = \mathcal{N}_{+}(\mu,0) + \mathcal{N}_{-}(\mu,0) = \sqrt{\mu}/\pi^2 $, $a = 0.1066$, $b = 0.0013$, and $c = 0.0064$ are numeric constants, and the functions $u_i(\alpha)$, $v_i(\alpha)$, and $w_i(\alpha)$ are computed to be
\begin{align*}
u_1(\alpha) &= -\frac{1}{420} (216 \sin 2 \alpha -60 \sin 4 \alpha -460 \cos 2 \alpha
\\&\quad\quad\quad\quad
+73 \cos 4 \alpha +447),
\\
v_1(\alpha) &= \frac{1}{420}  (-648 \sin 2 \alpha +180 \sin 4 \alpha +260 \cos 2 \alpha
\\&\quad\quad\quad
+133 \cos 4 \alpha -573),
\\
w_1(\alpha) &= \frac{1}{70} (-68 \cos 2 \alpha +11 \cos 4 \alpha +61),
\\
u_2(\alpha) &= \frac{1}{210}    (-200 \cos 2 \alpha +77 \cos 4 \alpha -48 \sin \alpha \cos ^3 \alpha +108 ),
\\
v_2(\alpha) &= -\frac{1}{210}    (40 \cos 2 \alpha -43 \cos 4 \alpha +144 \sin \alpha \cos ^3 \alpha +48 ),
\\
w_2(\alpha) &= \frac{1}{35} (18 \cos 2 \alpha - 6 \cos 4 \alpha - 11),
\\
u_3(\alpha) &= -v_3(\alpha) =  -\frac{4}{35}  (19 \cos 2 \alpha-16 )  \sin^2 \alpha,
\\
w_3(\alpha) &= 0.
\end{align*}

\begin{figure}[!t]
\centering
\includegraphics[scale = 0.45]{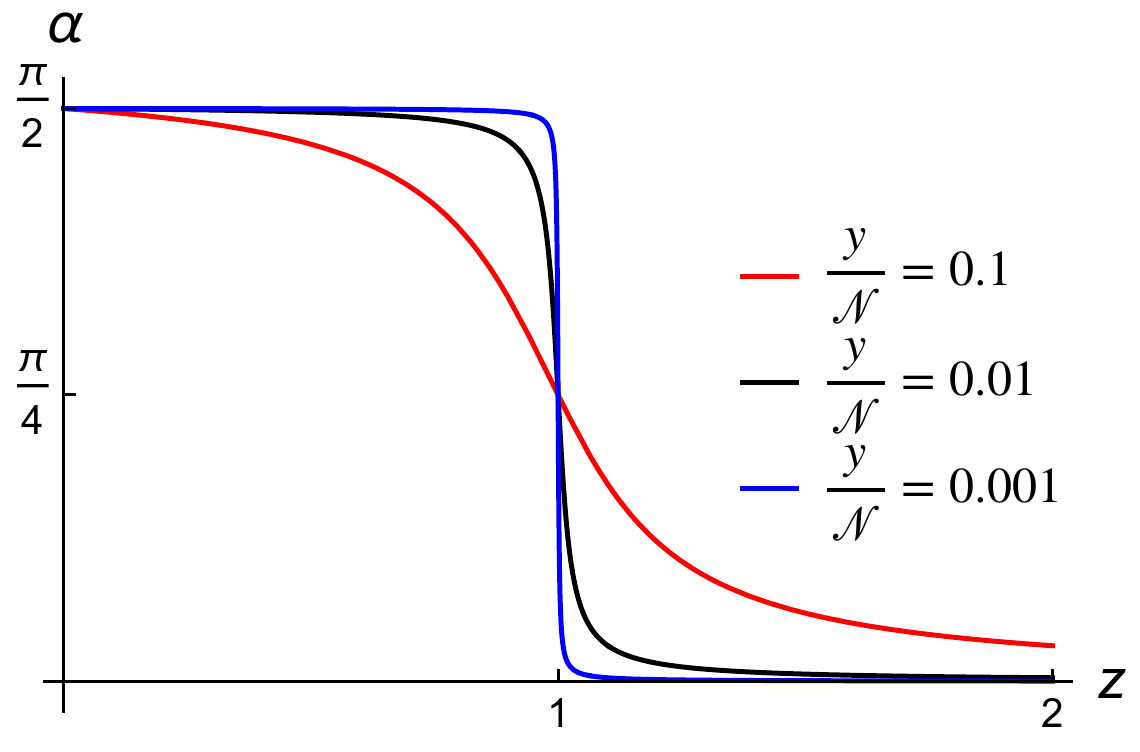}
\caption{(Color online) The superposition angle $\alpha$ as a function of the ratio $z=V_0/V_2$ given by Eq.~(\ref{cot_2}). We take three different values of $y/ \mathcal{N} =0.1$ (red), $y/\mathcal{N} =0.01$ (black), and $y/\mathcal{N} =0.001$ (blue). The range of $z$ for which the angle $\alpha$ deviates significantly from $0$ and $\pi/2$ broadens with increase of $y$.}
\label{fig_alpha_z}
\end{figure}

Let us begin by considering the limit $y\rightarrow 0$ when the spin-orbit coupling vanishes and consequently the superposition should reduce to either the orbital or the spin $d$-wave state. One thus expects that in this limit, for $V_0 > V_2$ it should be that $\alpha \to 0$, whereas for $V_2 > V_0$ one should find $\alpha \to \pi/2$. This corresponds to two orthonormal states $(1,0)^{\rm T}$ and $(0,1)^{\rm T}$, respectively. For $y \to 0$, Eq.~(\ref{eq_x0}) readily reduces to
\begin{align}
x_0 \simeq
\begin{cases}
-\dfrac{1}{\mathcal{N} V_0} + \mathcal{O}(y^2) \quad {\rm for} \quad V_0 > V_2,
\\ \\
-\dfrac{1}{\mathcal{N} V_2} + \mathcal{O}(y^2)  \quad  {\rm for}  \quad V_0 < V_2.
\end{cases}
\end{align}
Employing the last two relations and using Eq.~(\ref{eq_cot}) indeed leads to $\cot \alpha \to \infty$ and hence $\alpha \to 0$ for $ V_0 > V_2$, while $\cot \alpha \to 0 $ and thus $\alpha \to \pi/2$ for $ V_0 >  V_2$, in accordance with the expectation. By defining the ratio of the two interactions as
$z = V_0/V_2$, Eq.~(\ref{eq_cot}) can be rewritten as
\begin{align}
\alpha = {\rm Arccot} \Big( \frac{\big[\Lambda  - \mathcal{N}(z+1) \big]y}{\big[\Lambda  - \mathcal{N}(z-1) \big] \mathcal{N}  -2 y^2}
\Big),
\label{cot_2}
\end{align}
where $\Lambda = \sqrt{\mathcal{N}^2 (z-1)^2 + 4z y^2 }$. The full dependence of the angle $\alpha$ on  $z$ for various values of parameter $y$ is shown in Fig.~\ref{fig_alpha_z}. In the limit  of vanishing spin-orbit coupling $y \to 0$ the matrix $\mathcal{F}_{ab}(x)$ becomes diagonal and $\alpha$ approaches the step function. At infinitesimal $y$,  however, a full range of angles $\alpha$ is found by varying $z$ (see Fig.~\ref{fig_alpha_z}).

Let us consider the quartic coefficients next. First, take the inversion-symmetric case $\delta=0$, and consider the functions $w_i (\alpha)$.
The coefficient $q_1 >0$ for all angles, as we already announced. The function $w_2 (\alpha)$, on the other hand, changes sign at
\begin{align}
\alpha_{c,2} &= {\rm Arccos}(\dfrac{\sqrt{21-\sqrt{21}}}{2\sqrt{6}}) \simeq 0.60.
\end{align}

\begin{figure}[!t]
\centering
\includegraphics[scale = 0.3]{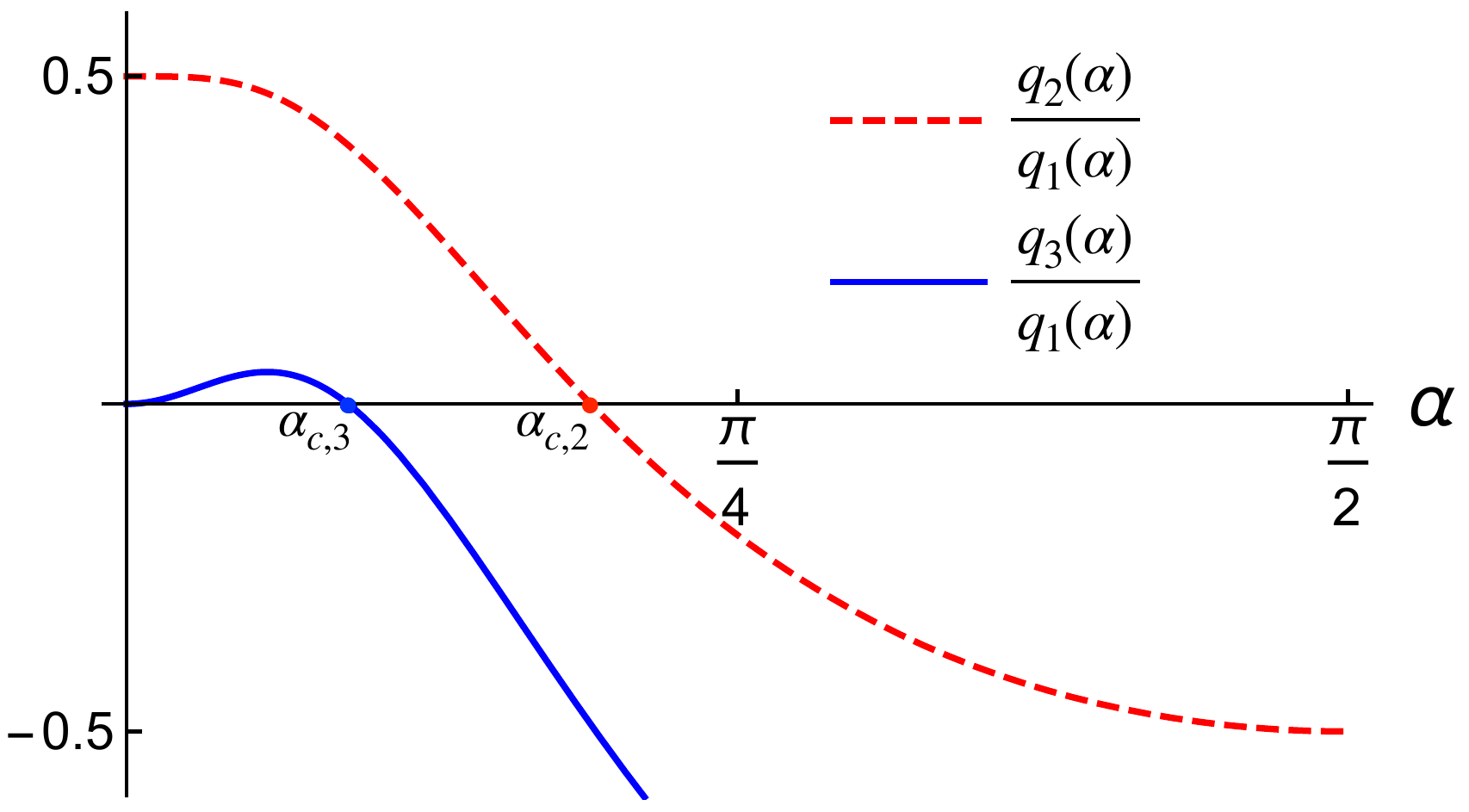}
\caption{(Color online) $q_2/q_1$ (dashed red line) and $q_3/q_1$ (solid blue line) as a function of the superposition angle $\alpha$. We set the parameter $\delta^2 \mu/T_c ^2 = 10 $ to make the curve $q_3/q_1$ visible on the scale of the figure.}
\label{fig2}
\end{figure}

The ratio of the coefficients $q_2/q_1$ is plotted as a function of the angle $\alpha$ on Fig. \ref{fig2}. The simple limits of $\alpha=0$ and $\alpha=\pi/2$ that we found agree with the previous studies. For $\alpha=0$, the $(l=2, s=0)$ pairing takes place on two Fermi surfaces separately, both of which lead to the ratio $q_2/q_1 = 1/2$ \cite{mandal}. For $\alpha=\pi/2$, the pairing is entirely in $(l=0, s=2)$ channel, which in the limit of small band-inversion yields $q_2/q_1 =-1/2$ \cite{majid1}. The smooth interpolation between these two limits as the superposition angle varies was therefore inevitable. This is our first result: The superconducting ground state shows a transition from real state to time-reversal symmetry broken state as a function of the ratio of the two pairing interactions is varied. In particular, for weak spin-orbit coupling almost the entire variation of the superposition angle happens in a narrow range of the ratio $z$ around the degenerate value $z=1$.

It is not surprising that for $\delta=0$ we find the values of the coefficient $q_3$ to be $q_3 (0) = q_3 (\pi/2)=0$, also in accordance with previous calculations \cite{mandal, majid1}.  Now we further find, however, that when $\delta=0$ the coefficient $q_3(\alpha) = 0$ for all intermediate values of $\alpha$ as well. A precise symmetry reason which could be behind this fact and the persisting accidental degeneracy between the ferromagnetic and cyclic phases which it causes (when $q_2 >0$) eludes us. Computationally, it follows after a lengthy calculation involving the Luttinger propagator in Eq. (\ref{prop_1}), which as a result of being a second-order in momentum and inversion-symmetric contains only the anticommuting gamma-matrices and the unit matrix. These matrices themselves can be expressed as particular even powers of $j=3/2$ angular momentum generators \cite{abrikosov, janssen}. Turning on a small perturbation kinetic energy term such as the one proportional to $\delta$, which breaks inversion by introducing odd powers of linear and angular momenta, however, does produce a finite coefficient $q_3$. In fact the coefficient $q_3$ at a small $\delta$ changes sign (see Fig. \ref{fig2}) at
\begin{align}
\alpha_{c,3} &= {\rm Arccos}(\sqrt{\dfrac{35}{38}}) \simeq 0.28.
\end{align}
This is our second result. Since $\alpha_{c,3} < \alpha_{c,2}$, the coefficient $q_3$ changes sign in the region where $q_2$ is positive, i.e., where the time-reversal symmetry is broken by the ground state. There is therefore an additional phase transition between the cyclic and the ferromagnetic phases as the interaction ratio $z$, and consequently the superposition angle $\alpha$, is changed. The trajectory through the phase diagram is depicted in Fig. \ref{fig3}.

\section{Summary and Discussion}

To summarize, if there is an attractive interaction in both the orbital $(l=2, s=0)$ and the spin $(l=0, s=2)$ $d$-wave channels, then, and as long as the spin-orbit coupling $\lambda \neq 0$, the resulting $d$-wave phase is a real linear superposition of the two states.  The superposition angle takes values between zero and $\pi/2$, and depends only on the ratio between the two attractive interactions. As first pointed out in the Ref. \cite{savary}, this was to be expected, and indeed similar superpositions are found in related problems \cite{link, yu}.

 \begin{figure}[!t]
\centering
\includegraphics[scale = 0.6]{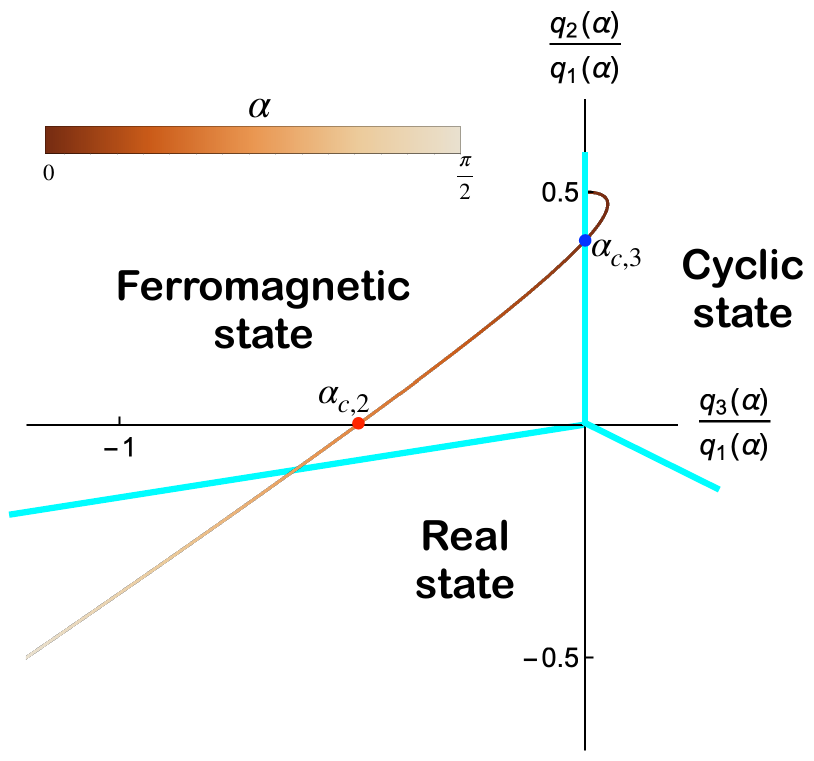}
\caption{(Color online) The curvy line represents the phases of the $d$-wave superconductor as the $z$-dependent superposition angle $\alpha$  is varied, and the cyan lines show the phase boundaries. When $q_3 = 0$, the ratio $q_2/q_1$ varies from 1/2 to -1/2 along the vertical axis. However, when the small inversion symmetry-breaking term makes $q_3$ finite, all three phases become available for different angles. We have again set $\delta^2 \mu/T_c^2 = 10$ for visual clarity. For weaker violation of inversion the system trajectory becomes closer to the $y$ axis.}
\label{fig3}
\end{figure}

A new feature that arises in this context is that once the quadratic term in the GL energy selects the preferred linear combination of the order parameters below $T_c$, the coefficients of the higher-order terms in principle depend on that particular linear combination, i.e., on the single ``superposition angle" in our example. This dependence was already observed in Ref. \cite{link} where a superposition of two $j=2$ states with $(l=1, s=1)$ and $(l=1, s=3)$, i.e., with $p$-wave orbital pairing was found to be the ground state configuration for a certain regime of parameters. In that case, however, the resulting $j=2$ state turned out to always be the same real state, irrespective of the linear combination. This is to be contrasted with the result found here. Namely, even when the inversion symmetry is present (i.e., when the parameter $\delta=0$) and consequently $q_3 (\alpha) = 0$, the coefficient $q_2 (\alpha)$ changes sign at the angle $\alpha_{c,2}$ as the ratio of the interactions is varied. This, although not sufficient by itself to fully determine the superconducting phase, already implies that the superconducting state has a transition from the one maximally breaking to the one fully preserving time-reversal symmetry as the ratio of the two interactions is varied. It is therefore particularly interesting that when a small inversion breaking term is included and $\delta \neq 0$, not only does one find that $q_3$ becomes finite but also that it changes sign at the superposition angle $\alpha_{c,3}$ which happens to be {\it smaller} than $\alpha_{c,2}$. This implies that as the ratio $z=V_0/ V_2 $ is varied from zero to infinity the system will find itself in all three phases of the superconducting phase diagram (Fig. \ref{fig3}). The low-energy density of states, as an example of an important observable known to be sensitive to violation of the time-reversal symmetry in multiband superconductors through the appearance of mini Bogoliubov-Fermi surfaces \cite{agterberg, link1, herbut1} may thus crucially depend on the value of a nonuniversal parameter such as $z$.

Although we considered the specific inversion-breaking kinetic energy term, it is clear that essentially any such term will yield a finite GL coefficient $q_3$. This is because the root cause for vanishing of the $q_3$ is the Clifford (anticommuting) algebra that is satisfied by the gamma matrices that appear exclusively in presence of inversion. Breaking of inversion introduces odd-powers of angular momentum generators in the Luttinger propagator, which do not satisfy Clifford algebra, and thus readily produce a finite $q_3$ \cite{comment1}. It is not guaranteed, however, that $q_3$ has to change sign with the variation of the interaction ratio $z$. This result may not be generic. But if it were not, as we suspect it might not be, since the change of sign of the coefficient $q_2$ is generic, then the system would still exhibit at least two different phases as $z$ is varied. It is also interesting that with our choice of the inversion-breaking term, we find cyclic phase become a possible ground state, in contrast to Ref. \cite{ho}, where, as the spin of paired particles is varied from 3/2 to infinity, the only transition is between the real and the ferromagnetic phases.

\section*{Acknowledgement}
This work has been supported by the NSERC of Canada.

\begin{appendix}
\setcounter{figure}{0}
\renewcommand{\thefigure}{C\arabic{figure}}
\section{The quadratic coefficients of the GL}
\label{app_A}
In this Appendix, the explicit form of the matrix $\mathcal{F}_{ab}$ in Eq.~(\ref{F_matrix}) is calculated. First, it can easily be shown that
\begin{align}
\mathcal{F}_{ab} =
\begin{pmatrix}
\dfrac{\delta_{ab}}{V_0} + K^{0,0}_{ab}
&&
K^{0,2}_{ab}
\\ \\
K^{2,0}_{ab}
&&
\dfrac{\delta_{ab}}{V_2} + K^{2,2}_{ab}
\end{pmatrix},
\end{align}
and
\begin{equation*}
K^{s,s'}_{ab} = -2T \sum_n
\int \frac{d^3 \bm{p}}{(2\pi)^3}
{\rm tr} \Big[
G_0(-Q) \mathcal{A}^{(s) }_{a} G_0(Q) \mathcal{A}^{(s')}_b
\Big].
\end{equation*}
The loop integrals in the diagonal components of the matrix $\mathcal{F}_{ab}$ when inversion symmetry breaking parameter $\delta=0$ are given by
\begin{widetext}
\begin{align}
K^{0,0}_{ab} &=  - \frac{8T \delta_{ab} }{5} \sum_n
\int \frac{d^3 \bm{p}}{(2\pi)^3}
\frac{
p^4 (1 + \lambda^2)-2 p^2 \mu +\mu ^2 +\omega_n^2
}{\Big[
(p^2(1 + \lambda)-\mu)^2 + \omega_n^2
\Big]\Big[
(p^2(1 - \lambda)-\mu)^2 + \omega_n^2
\Big]},
\\
K^{2,2}_{ab} &=  - 8T \delta_{ab} \sum_n
\int \frac{d^3 \bm{p}}{(2\pi)^3}
\frac{
p^4 (1-\frac{3}{5}\lambda^2)-2 p^2  \mu +\mu ^2 +\omega_n ^2
}{\Big[
(p^2(1 + \lambda)-\mu)^2 + \omega_n^2
\Big]\Big[
(p^2(1 - \lambda)-\mu)^2 + \omega_n^2
\Big]}.
\end{align}
The integrands appearing in these expressions can be better written as
\begin{align}
\frac{p^4 (1 + r \lambda^2)-2p^2 \mu + \mu ^2 + \omega_n ^2
}{(\xi_{+}^2 + \omega_n^2)
(\xi_{-}^2 + \omega_n^2)}
&=
-\frac{1}{4}
\Big(
\frac{\xi_{+} + \xi_{-}}{\xi_{+} - \xi_{-}}
+r\frac{\xi_{+} - \xi_{-}}{\xi_{+} + \xi_{-}}
+\frac{4\omega_n^2}{\xi_{+}^2 - \xi_{-}^2}
\Big)
\Bigg[
\frac{1}{\xi_{+}^2 + \omega_n^2} -
\frac{1}{\xi_{-}^2 + \omega_n^2}
\Bigg]
\nonumber
\\&=
\frac{1}{4}
\sum_{\eta = \pm}
\frac{1}{\xi_{\eta}^2 + \omega_n^2}
\Big[
1+r + 2 (1-r) \frac{\xi_{\eta} }{\xi_{-\eta}} + \frac{4 \omega_n ^2}{\xi_{-\eta} } + O(\xi_{\eta} ^2 )
\Big] ,
\end{align}
\end{widetext}
where $\xi_{\pm}(p) = p^2(1 \pm \lambda)-\mu$ and $r=1 (r=-3/5)$ for $K^{0,0}_{ab} (K^{2,2}_{ab})$ component. Therefore, to the leading order at low temperatures
\begin{align}
K^{0,0}_{ab}  &=  -
\frac{2 T \delta_{ab} }{5} \mathcal{N}
\sum_{n}
\int_{-\Omega}^{\Omega}
\frac{d \xi}{\xi^2 + \omega_n^2}
\nonumber
\\&=
-\frac{2 \delta_{ab} }{5}   \mathcal{N}
\Big[
 \log(\frac{\Omega}{T}) + {\rm const}
 \Big] ,
\end{align}
where we used
\begin{align*}
p^2 -\mu &= \frac{\xi_{+} + \xi_{-} }{2},
\\
\lambda p^2  &= \frac{\xi_{+} - \xi_{-} }{2},
\\
\int \frac{d^3 \bm{p}}{(2\pi)^3} =
\frac{\mathcal{N}_{+}(\mu,\lambda)}{2}
& \int d\xi_{+}
=
\frac{\mathcal{N}_{-}(\mu,\lambda)}{2}
\int d\xi_{-},
\end{align*}
and the definitions of $\mathcal{N}$, $\mathcal{N}_{+}(\mu,\lambda)$, and $\mathcal{N}_{-}(\mu,\lambda)$ are given in the main text. Similarly, to the leading order at low temperatures one also finds
\begin{align}
K^{2,2}_{ab} = - \frac{2 \delta_{ab} }{5}    \mathcal{N}
\Big[
\log(\frac{\Omega}{T})+ {\rm const}
\Big] .
\end{align}
For the off-diagonal components, we have
\begin{widetext}
\begin{align}
K^{1,2}_{ab} = K^{2,1}_{ab} &= \frac{16T \delta_{ab} }{5}  \sum_n
\int \frac{d^3 \bm{p}}{(2\pi)^3}
\frac{p^2 \lambda (p^2  -\mu)}{\Big[
(p^2(1 + \lambda)-\mu)^2 + \omega_n^2
\Big]\Big[
(p^2(1 - \lambda)-\mu)^2 + \omega_n^2
\Big]}
\nonumber
\\&
= \frac{16T \delta_{ab} }{5}  \sum_n
\int \frac{d\xi_{\eta}}{2}
\frac{\xi_{+}^2 - \xi_{-}^2}{
4(\xi_{-}^2 - \xi_{+}^2)}
\Big[
\frac{\mathcal{N}_{+}}{(\xi_{+}^2 + \omega_n^2)}-
\frac{\mathcal{N}_{-}}{(\xi_{-}^2 + \omega_n^2)}
\Big]
\nonumber
\\&
= \frac{2 \delta_{ab} }{5}
(\mathcal{N}_{-}-\mathcal{N}_{+})
\Big[
 \log(\frac{\Omega}{T}) + {\rm const}
\Big].
\end{align}
\end{widetext}
By introducing $x = -\frac{2}{5}\log(\frac{\Omega}{T})$ and $y = \mathcal{N}_{-} - \mathcal{N}_{+}$, we obtain the matrix form shown in Eq.~(\ref{F_matrix}) in the main text.
\section{The quartic coefficients of the GL}
\label{app_B}
From one-loop integral, we have
\begin{align}
F_4(\Vec{\Delta}) =
\sum_{a,b,c,d=1}^5
K_{abcd}
\Delta_a \Delta^{\ast}_b
\Delta_c \Delta^{\ast}_d,
\end{align}
where
\begin{widetext}
\begin{align}
K_{abcd} = 4T \sum_{n=-\infty}^{\infty}
\int \frac{d^3 \bm{p}}{(2\pi)^3}
{\rm tr} \big[
G(-Q) \mathcal{B}_{a}(\bm{p},\alpha) G(Q) \mathcal{B}_b(\bm{p},\alpha)
G(-Q) \mathcal{B}_{c}(\bm{p},\alpha) G(Q) \mathcal{B}_d(\bm{p},\alpha)
\big],
\end{align}
\end{widetext}
for
 \begin{align}
G(Q) =
\frac{(i \omega_n  - p^2 + \mu) \mathds{1}_{4 \times 4} + \lambda d_a(\bm{p}) \gamma_a + \delta \bm{J} \cdot \bm{p} }{(i \omega_n  - p^2 + \mu)^2 - \lambda^2 p^4},
\nonumber
\end{align}
 and
\begin{align}
\mathcal{B}_a(\bm{p},\alpha) &= \frac{d_a(\bm{p})}{p^2} \mathds{1}_{4 \times 4} \cos \alpha  +  \gamma_a \sin \alpha,
\nonumber
\end{align}
are five $4 \times 4$ Hermitian matrices where $a=1,2,\dots,5$. Here, $\bm{J} = (J_x, J_y, J_z)$ are the usual spin-$\frac{3}{2}$ generators of the SO(3)
\begin{align*}
J_x &= \frac{1}{2}
\begin{pmatrix}
0   & \sqrt{3}  & 0   &  0 \\
\sqrt{3} & 0 & 2 & 0 \\
0 & 2&0&\sqrt{3} \\
0&0&\sqrt{3}&0
\end{pmatrix},
\\
J_y =& \frac{i}{2}
\begin{pmatrix}
0   & -\sqrt{3}  & 0   &  0 \\
\sqrt{3} & 0 & -2 & 0 \\
0 & 2 & 0 & -\sqrt{3} \\
0& 0 & \sqrt{3} & 0
\end{pmatrix},
\\
J_z &= \frac{1}{2}
\begin{pmatrix}
3 & 0  & 0   &  0 \\
0& 1 & 0 & 0 \\
0& 0 & -1 & 0 \\
0& 0 & 0 & -3 \\
\end{pmatrix}.
\end{align*}
In order to calculate the coefficients $q_{1,2,3}$, we substitute the distinct states  $\Vec{\Delta}_1 = (1,0,0,0,0)$, $\Vec{\Delta}_2 = (\frac{1}{\sqrt{2}},\frac{i}{\sqrt{2}},0,0,0)$, $\Vec{\Delta}_3 = (\frac{1}{\sqrt{2}},0,\frac{i}{\sqrt{2}},0,0)$, and match them with
\begin{align}
F_4(\Vec{\Delta}_1) &= 4 (q_1 + q_2) + 2 q_3,
\\
F_4(\Vec{\Delta}_2) &= 4 q_1 + \frac{4}{3}q_3,
\\
F_4(\Vec{\Delta}_3) &= 4 q_1 + 2 q_3,
\end{align}
where we used Eq.~(\ref{eq_quartic}) in the main text. Solving these three linearly independent equations simultaneously yields
\begin{widetext}
\begin{align}
q_i &= 4T \sum_{\omega_n} \int \frac{d^3 \bm{p}}{(2\pi)^3}
\frac{f_i(p,\omega_n,\lambda,\delta,\mu,\alpha)}{
\Big[(p^2(1+ \lambda) - \mu)^2 + \omega^2_n \Big]^2
\Big[(p^2(1 - \lambda) - \mu)^2 + \omega^2_n \Big]^2
},
\end{align}
where
\begin{align*}
f_1(p,\omega_n,\lambda,\delta,\mu,\alpha)  &=
\frac{3}{10} \lambda ^4 p^8-\frac{2}{5} \lambda ^4 p^8 \cos 2 \alpha  +\frac{48}{35} \lambda ^2 p^8 \cos 2 \alpha   - \frac{34}{35} p^8 \cos 2 \alpha  +\frac{11}{70} \lambda ^4 p^8 \cos 4 \alpha  -\frac{19}{35} \lambda ^2 p^8 \cos 4 \alpha
\\& \quad
 +\frac{11}{70} p^8 \cos 4 \alpha  +\frac{4}{35} \lambda ^3 p^8 \sin 2 \alpha  -\frac{36}{35} \lambda  p^8 \sin 2 \alpha  -\frac{2}{7} \lambda ^3 p^8 \sin 4 \alpha  +\frac{2}{7} \lambda  p^8 \sin 4 \alpha  -\frac{17}{35} \lambda ^2 p^8
 \\& \quad
 +\frac{61 p^8}{70}+\frac{1}{4} \delta ^2 \lambda ^2 p^6-\frac{17}{35} \delta ^2 \lambda  p^6+\frac{34}{35} \lambda ^2 \mu  p^6+\frac{23}{21} \delta ^2 p^6 \cos 2 \alpha  -\frac{44}{105} \delta ^2 \lambda ^2 p^6 \cos 2 \alpha  +\frac{124}{105} \delta ^2 \lambda  p^6 \cos 2 \alpha
 \\& \quad
 -\frac{96}{35} \lambda ^2 \mu  p^6 \cos 2 \alpha  +\frac{136}{35} \mu  p^6 \cos 2 \alpha  -\frac{73}{420} \delta ^2 p^6 \cos 4 \alpha  +\frac{11}{420} \delta ^2 \lambda ^2 p^6 \cos 4 \alpha  -\frac{7}{15} \delta ^2 \lambda  p^6 \cos 4 \alpha
 \\& \quad
 +\frac{38}{35} \lambda ^2 \mu  p^6 \cos 4 \alpha  -\frac{22}{35} \mu  p^6 \cos 4 \alpha  -\frac{18}{35} \delta ^2 p^6 \sin 2 \alpha  +\frac{2}{35} \delta ^2 \lambda ^2 p^6 \sin 2 \alpha  +\frac{13}{35} \delta ^2 \lambda  p^6 \sin 2 \alpha
 \\& \quad
 -\frac{4}{35} \lambda ^3 \mu  p^6 \sin 2 \alpha  +\frac{108}{35} \lambda  \mu  p^6 \sin 2 \alpha  +\frac{1}{7} \delta ^2 p^6 \sin 4 \alpha  -\frac{1}{7} \delta ^2 \lambda ^2 p^6 \sin 4 \alpha  +\frac{1}{10} \delta ^2 \lambda  p^6 \sin 4 \alpha
 \\& \quad
 +\frac{2}{7} \lambda ^3 \mu  p^6 \sin 4 \alpha  -\frac{6}{7} \lambda  \mu  p^6 \sin 4 \alpha  -\frac{122}{35} \mu  p^6-\frac{149}{140} \delta ^2 p^6+\frac{109}{1120} \delta ^4 p^4 -\frac{17}{35} \lambda ^2 \mu ^2 p^4+\frac{183}{35} \mu ^2 p^4
 \\& \quad
 -\frac{13}{5} \lambda ^2 \omega_n^2 p^4+\frac{61}{35} \omega_n^2 p^4+\frac{149}{70} \delta ^2 \mu  p^4+\frac{17}{35} \delta ^2 \lambda  \mu  p^4+\frac{31}{280} \delta ^4 p^4 \cos 2 \alpha   +\frac{48}{35} \lambda ^2 \mu ^2 p^4 \cos 2 \alpha
  \\& \quad
-\frac{204}{35} \mu ^2 p^4 \cos 2 \alpha  +\frac{128}{35} \lambda ^2 \omega_n^2 p^4 \cos 2 \alpha  -\frac{68}{35} \omega_n^2 p^4 \cos 2 \alpha  -\frac{46}{21} \delta ^2 \mu  p^4 \cos 2 \alpha  -\frac{124}{105} \delta ^2 \lambda  \mu  p^4 \cos 2 \alpha
 \\& \quad
-\frac{19}{35} \lambda ^2 \mu ^2 p^4 \cos 4 \alpha  +\frac{33}{35} \mu ^2 p^4 \cos 4 \alpha  -\frac{33}{35} \lambda ^2 \omega_n^2 p^4 \cos 4 \alpha  +\frac{11}{35} \omega_n^2 p^4 \cos 4 \alpha  +\frac{73}{210} \delta ^2 \mu  p^4 \cos 4 \alpha
 \\& \quad
+\frac{7}{15} \delta ^2 \lambda  \mu  p^4 \cos 4 \alpha  +\frac{13}{70} \delta ^4 p^4 \sin 2 \alpha  -\frac{108}{35} \lambda  \mu ^2 p^4 \sin 2 \alpha  -\frac{36}{35} \lambda  \omega_n^2 p^4 \sin 2 \alpha  +\frac{36}{35} \delta ^2 \mu  p^4 \sin 2 \alpha
 \\& \quad
-\frac{13}{35} \delta ^2 \lambda  \mu  p^4 \sin 2 \alpha  +\frac{1}{20} \delta ^4 p^4 \sin 4 \alpha  +\frac{6}{7} \lambda  \mu ^2 p^4 \sin 4 \alpha  +\frac{2}{7} \lambda  \omega_n^2 p^4 \sin 4 \alpha  -\frac{2}{7} \delta ^2 \mu  p^4 \sin 4 \alpha
 \\& \quad
-\frac{1}{10} \delta ^2 \lambda  \mu  p^4 \sin 4 \alpha  -\frac{69}{1120}\delta ^4 p^4 \cos 4 \alpha -\frac{149}{140} \delta ^2 \mu ^2 p^2-\frac{191}{140} \delta ^2 \omega_n^2 p^2-\frac{122}{35} \mu  \omega_n^2 p^2+\frac{136}{35} \mu ^3 p^2 \cos 2 \alpha
\\& \quad
+\frac{23}{21} \delta ^2 \mu ^2 p^2 \cos 2 \alpha  +\frac{13}{21} \delta ^2 \omega_n^2 p^2 \cos 2 \alpha  +\frac{136}{35} \mu  \omega_n^2 p^2 \cos 2 \alpha  -\frac{22}{35} \mu ^3 p^2 \cos 4 \alpha  -\frac{73}{420} \delta ^2 \mu ^2 p^2 \cos 4 \alpha
\\& \quad
+\frac{19}{60} \delta ^2 \omega_n^2 p^2 \cos 4 \alpha  -\frac{22}{35} \mu  \omega_n^2 p^2 \cos 4 \alpha +\frac{36}{35} \lambda  \mu ^3 p^2 \sin 2 \alpha  -\frac{18}{35} \delta ^2 \mu ^2 p^2 \sin 2 \alpha  -\frac{54}{35} \delta ^2 \omega_n^2 p^2 \sin 2 \alpha
\\& \quad
+\frac{36}{35} \lambda  \mu  \omega_n^2 p^2 \sin 2 \alpha  -\frac{2}{7} \lambda  \mu ^3 p^2 \sin 4 \alpha  +\frac{1}{7} \delta ^2 \mu ^2 p^2 \sin 4 \alpha  +\frac{3}{7} \delta ^2 \omega_n^2 p^2 \sin 4 \alpha  -\frac{2}{7} \lambda  \mu  \omega_n^2 p^2 \sin 4 \alpha
\\& \quad
-\frac{122}{35} \mu ^3 p^2+\frac{61}{70} \mu ^4 +\frac{61}{70}\omega_n^4+\frac{61}{35} \mu ^2 \omega_n^2 -\frac{34}{35} \mu ^4 \cos 2 \alpha -\frac{34}{35} \omega_n^4 \cos 2 \alpha -\frac{68}{35} \mu ^2 \omega_n^2 \cos 2 \alpha
 \\& \quad
+\frac{11}{70} \mu ^4 \cos 4 \alpha +\frac{11}{70} \omega_n^4 \cos 4 \alpha +\frac{11}{35} \mu ^2 \omega_n^2 \cos 4 \alpha ,
\end{align*}
\begin{align*}
f_2(p,\omega_n,\lambda,\delta,\mu,\alpha)  &=
\frac{6}{7}  \lambda ^2 p^8+\frac{2}{5} \lambda ^4 p^8 \cos 2 \alpha  -\frac{32}{35} \lambda ^2 p^8 \cos 2 \alpha  +\frac{18}{35} p^8 \cos 2 \alpha
-\frac{6}{35} \lambda ^4 p^8 \cos 4 \alpha  +\frac{8}{35} \lambda ^2 p^8 \cos 4 \alpha
 \\& \quad
-\frac{6}{35} p^8 \cos 4 \alpha  -\frac{12}{35} \lambda ^3 p^8 \sin 2 \alpha  -\frac{4}{35} \lambda  p^8 \sin 2 \alpha  +\frac{2}{35} \lambda ^3 p^8 \sin 4 \alpha  -\frac{2}{35} \lambda  p^8 \sin 4 \alpha  -\frac{1}{5}\lambda ^4 p^8-\frac{11}{35} p^8
 \\& \quad
+\frac{18}{35} \delta ^2 p^6 + \frac{2}{7} \delta ^2 \lambda ^2 p^6+\frac{37}{35} \delta ^2 \lambda  p^6-\frac{12}{7} \lambda ^2 \mu  p^6+\frac{44 }{35}\mu  p^6-\frac{20}{21} \delta ^2 p^6 \cos 2 \alpha  -\frac{73}{105} \delta ^2 \lambda ^2 p^6 \cos 2 \alpha
 \\& \quad
-\frac{148}{105} \delta ^2 \lambda  p^6 \cos 2 \alpha  +\frac{64}{35} \lambda ^2 \mu  p^6 \cos 2 \alpha  -\frac{72}{35} \mu  p^6 \cos 2 \alpha  +\frac{11}{30} \delta ^2 p^6 \cos 4 \alpha  +\frac{71}{210} \delta ^2 \lambda ^2 p^6 \cos 4 \alpha
 \\& \quad
+\frac{7}{15} \delta ^2 \lambda  p^6 \cos 4 \alpha  -\frac{16}{35} \lambda ^2 \mu  p^6 \cos 4 \alpha  +\frac{24}{35} \mu  p^6 \cos 4 \alpha  -\frac{2}{35} \delta ^2 p^6 \sin 2 \alpha  -\frac{6}{35} \delta ^2 \lambda ^2 p^6 \sin 2 \alpha
 \\& \quad
+\frac{11}{35} \delta ^2 \lambda  p^6 \sin 2 \alpha  +\frac{12}{35} \lambda ^3 \mu  p^6 \sin 2 \alpha  +\frac{12}{35} \lambda  \mu  p^6 \sin 2 \alpha  -\frac{1}{35} \delta ^2 p^6 \sin 4 \alpha  +\frac{1}{35} \delta ^2 \lambda ^2 p^6 \sin 4 \alpha
 \\& \quad
-\frac{1}{70} \delta ^2 \lambda  p^6 \sin 4 \alpha  -\frac{2}{35} \lambda ^3 \mu  p^6 \sin 4 \alpha  +\frac{6}{35} \lambda  \mu  p^6 \sin 4 \alpha  +\frac{\delta ^4 p^4}{80}+\frac{6}{7} \lambda ^2 \mu ^2 p^4+\frac{6}{5} \lambda ^2 \omega_n^2 p^4-\frac{36}{35} \delta ^2 \mu  p^4
 \\& \quad
-\frac{37}{35} \delta ^2 \lambda  \mu  p^4+\frac{47}{280} \delta ^4 p^4 \cos 2 \alpha  -\frac{32}{35} \lambda ^2 \mu ^2 p^4 \cos 2 \alpha  +\frac{108}{35} \mu ^2 p^4 \cos 2 \alpha  -\frac{48}{35} \lambda ^2 \omega_n^2 p^4 \cos 2 \alpha
 \\& \quad
+\frac{36}{35} \omega_n^2 p^4 \cos 2 \alpha  +\frac{40}{21} \delta ^2 \mu  p^4 \cos 2 \alpha  +\frac{148}{105} \delta ^2 \lambda  \mu  p^4 \cos 2 \alpha  -\frac{3}{28} \delta ^4 p^4 \cos 4 \alpha  +\frac{8}{35} \lambda ^2 \mu ^2 p^4 \cos 4 \alpha
 \\& \quad
-\frac{36}{35} \mu ^2 p^4 \cos 4 \alpha  +\frac{8}{35} \lambda ^2 \omega_n^2 p^4 \cos 4 \alpha  -\frac{12}{35} \omega_n^2 p^4 \cos 4 \alpha  -\frac{11}{15} \delta ^2 \mu  p^4 \cos 4 \alpha  -\frac{7}{15} \delta ^2 \lambda  \mu  p^4 \cos 4 \alpha
 \\& \quad
 +\frac{11}{70} \delta ^4 p^4 \sin 2 \alpha  -\frac{12}{35} \lambda  \mu ^2 p^4 \sin 2 \alpha  -\frac{4}{35} \lambda  \omega_n^2 p^4 \sin 2 \alpha  +\frac{4}{35} \delta ^2 \mu  p^4 \sin 2 \alpha  -\frac{11}{35} \delta ^2 \lambda  \mu  p^4 \sin 2 \alpha
  \\& \quad
 -\frac{1}{140} \delta ^4 p^4 \sin 4 \alpha  -\frac{6}{35} \lambda  \mu ^2 p^4 \sin 4 \alpha  -\frac{2}{35} \lambda  \omega_n^2 p^4 \sin 4 \alpha  +\frac{2}{35} \delta ^2 \mu  p^4 \sin 4 \alpha  +\frac{1}{70} \delta ^2 \lambda  \mu  p^4 \sin 4 \alpha
  \\& \quad
 -\frac{66}{35} \mu ^2 p^4 -\frac{22}{35} \omega_n^2 p^4 +\frac{44 }{35} \mu ^3 p^2+\frac{18}{35} \delta ^2 \mu ^2 p^2-\frac{8}{35} \delta ^2 \omega_n^2 p^2+\frac{44}{35} \mu  \omega_n^2 p^2-\frac{72}{35} \mu ^3 p^2 \cos 2 \alpha   \\& \quad
 -\frac{20}{21} \delta ^2 \mu ^2 p^2 \cos 2 \alpha  -\frac{4}{21} \delta ^2 \omega_n^2 p^2 \cos 2 \alpha  -\frac{72}{35} \mu  \omega_n^2 p^2 \cos 2 \alpha  +\frac{24}{35} \mu ^3 p^2 \cos 4 \alpha  +\frac{11}{30} \delta ^2 \mu ^2 p^2 \cos 4 \alpha   \\& \quad
 +\frac{43}{210} \delta ^2 \omega_n^2 p^2 \cos 4 \alpha  +\frac{24}{35} \mu  \omega_n^2 p^2 \cos 4 \alpha  +\frac{4}{35} \lambda  \mu ^3 p^2 \sin 2 \alpha  -\frac{2}{35} \delta ^2 \mu ^2 p^2 \sin 2 \alpha  -\frac{6}{35} \delta ^2 \omega_n^2 p^2 \sin 2 \alpha
  \\& \quad
 +\frac{4}{35} \lambda  \mu  \omega_n^2  p^2 \sin 2 \alpha  +\frac{2}{35} \lambda  \mu ^3 p^2 \sin 4 \alpha  -\frac{1}{35} \delta ^2 \mu ^2 p^2 \sin 4 \alpha  -\frac{3}{35} \delta ^2 \omega_n^2 p^2 \sin 4 \alpha +\frac{2}{35} \lambda  \mu  \omega_n^2 p^2 \sin 4 \alpha
 \\& \quad
 +\frac{18}{35} \mu ^4 \cos 2 \alpha +\frac{18}{35} \omega_n^4 \cos 2 \alpha +\frac{36}{35} \mu ^2 \omega_n^2 \cos 2 \alpha -\frac{6}{35} \mu ^4 \cos 4 \alpha -\frac{6}{35} \omega_n^4 \cos 4 \alpha -\frac{12}{35} \mu ^2 \omega_n^2 \cos 4 \alpha
  \\& \quad
 -\frac{11}{35}\mu ^4-\frac{11}{35} \omega_n^4-\frac{22}{35} \mu ^2 \omega_n^2,
\end{align*}
\begin{align*}
f_3(p,\omega_n,\lambda,\delta,\mu,\alpha)  &=
\frac{16}{35} \delta ^2 \lambda ^2 p^6 \sin^2 \alpha -\frac{4}{5} \delta ^2 \lambda ^2 p^6 \sin^2 \alpha  \cos 2 \alpha +\frac{16}{5} \delta ^2 \lambda  p^6 \sin^2 \alpha +\frac{24}{35} \delta ^2 \lambda  p^6 \sin^2 \alpha  \sin 2 \alpha
 \\& \quad
-\frac{16}{5} \delta ^2 \lambda  p^6 \sin^2 \alpha  \cos 2 \alpha +\frac{64}{35} \delta ^2 p^6 \sin^2 \alpha -\frac{76}{35} \delta ^2 p^6 \sin^2 \alpha  \cos 2 \alpha +\frac{18}{35} \delta ^4 p^4 \sin^2 \alpha
 \\& \quad
+\frac{12}{35} \delta ^4 p^4 \sin^2 \alpha  \sin 2 \alpha -\frac{3}{35} \delta ^4 p^4 \sin^2 \alpha  \cos 2 \alpha -\frac{16}{5} \delta ^2 \lambda  \mu  p^4 \sin^2 \alpha -\frac{24}{35} \delta ^2 \lambda  \mu  p^4 \sin^2 \alpha  \sin 2 \alpha
 \\& \quad
+\frac{16}{5} \delta ^2 \lambda  \mu  p^4 \sin^2 \alpha  \cos 2 \alpha -\frac{128}{35} \delta ^2 \mu  p^4 \sin^2 \alpha +\frac{152}{35} \delta ^2 \mu  p^4 \sin^2 \alpha  \cos 2 \alpha +\frac{64}{35} \delta ^2 \mu ^2 p^2 \sin^2 \alpha
 \\& \quad
-\frac{76}{35} \delta ^2 \mu ^2 p^2 \sin^2 \alpha  \cos 2 \alpha -\frac{64}{35} \delta ^2 p^2 \omega_n^2 \sin^2 \alpha +\frac{76}{35} \delta ^2 p^2 \omega_n^2 \sin^2 \alpha  \cos 2 \alpha .
\end{align*}

For small $\lambda$, the integrands can be expanded as
\begin{align*}
q_i = 4T \sum_{n} \int \frac{d^3 p}{(2\pi)^3}
\frac{1}{{[(p^2 - \mu)^2 + \omega^2_n}]^4}
\Big[
h_i(p,\omega_n,\delta,\mu,\alpha) +  \mathcal{O}(\lambda)
\Big],
\end{align*}
where $h_i(p,\omega_n,\delta,\mu,\alpha) =
a_i (p,\omega_n,\mu,\alpha)  \delta^4
+ b_i (p,\omega_n,\mu,\alpha) \delta^2
+ c_i (p,\omega_n,\mu,\alpha) $ and
\begin{align*}
a_1(p,\omega_n,\mu,\alpha) &= \frac{p^4}{1120} (208 \sin 2 \alpha +56 \sin 4 \alpha +124 \cos 2 \alpha -69 \cos 4 \alpha +109),
\\
b_1(p,\omega_n,\mu,\alpha) &= -\frac{18}{35} p^6 \sin 2 \alpha +\frac{1}{7} p^6 \sin 4 \alpha +\frac{23}{21} p^6 \cos 2 \alpha -\frac{73}{420} p^6 \cos 4 \alpha -\frac{149}{140} p^6+\frac{36}{35} \mu  p^4 \sin 2 \alpha
\\&\quad
-\frac{2}{7} \mu  p^4 \sin 4 \alpha
-\frac{46}{21} \mu  p^4 \cos 2 \alpha +\frac{73}{210} \mu  p^4 \cos 4 \alpha +\frac{149}{70}\mu  p^4-\frac{18}{35} \mu ^2 p^2 \sin 2 \alpha +\frac{1}{7} \mu ^2 p^2 \sin 4 \alpha
\\&\quad
+\frac{23}{21} \mu ^2 p^2 \cos 2 \alpha
-\frac{73}{420} \mu ^2 p^2 \cos 4 \alpha -\frac{54}{35} p^2 \omega_n^2 \sin 2 \alpha +\frac{3}{7} p^2 \omega_n^2 \sin 4 \alpha +\frac{13}{21} p^2 \omega_n^2 \cos 2 \alpha +\frac{19}{60} p^2 \omega_n^2 \cos 4 \alpha
\\&\quad
-\frac{149}{140} \mu ^2 p^2
-\frac{191}{140} p^2 \omega_n^2,
\\
c_1(p,\omega_n,\mu,\alpha) &=  \frac{1}{70} \big[ (p^2-\mu  )^2+\omega_n^2 \big]^2(-68 \cos 2 \alpha +11 \cos 4 \alpha +61),
\\
a_2(p,\omega_n,\mu,\alpha) &= \frac{p^4}{560}  (88 \sin 2 \alpha -4 \sin 4 \alpha +94 \cos 2 \alpha -60 \cos 4 \alpha +7),
\\
b_2(p,\omega_n,\mu,\alpha) &= -\frac{20}{21} p^6 \cos 2 \alpha +\frac{11}{30} p^6 \cos 4 \alpha -\frac{8}{35} p^6 \sin \alpha  \cos ^3 \alpha +\frac{18 p^6}{35}+\frac{40}{21} \mu  p^4 \cos 2 \alpha -\frac{11}{15} \mu  p^4 \cos 4 \alpha
\\&\quad
+\frac{16}{35} \mu  p^4 \sin \alpha  \cos ^3 \alpha -\frac{36}{35}\mu  p^4-\frac{20}{21} \mu ^2 p^2 \cos 2 \alpha +\frac{11}{30} \mu ^2 p^2 \cos 4 \alpha -\frac{8}{35} \mu ^2 p^2 \sin \alpha  \cos ^3 \alpha
\\&\quad
-\frac{4}{21} p^2 \omega_n^2 \cos 2 \alpha +\frac{43}{210} p^2 \omega_n^2 \cos 4 \alpha -\frac{24}{35} p^2 \omega_n^2 \sin \alpha  \cos ^3 \alpha +\frac{18 }{35} \mu ^2 p^2 -\frac{8 }{35}p^2 \omega_n^2,
\\
c_2(p,\omega_n,\mu,\alpha) &=  -\frac{1}{35} \big[ (p^2-\mu  )^2+\omega_n^2  \big]^2 (-18 \cos 2 \alpha +6 \cos 4 \alpha +11),
\\
a_3(p,\omega_n,\mu,\alpha) &= \frac{3}{35} p^4  (4 \sin 2 \alpha -\cos 2 \alpha+ 6) \sin ^2 \alpha,
\\
b_3(p,\omega_n,\mu,\alpha) &= -\frac{4}{35}  p^2 \big[ (p^2-\mu  )^2-\omega_n^2 \big] (19 \cos 2 \alpha -16)   \sin ^2 \alpha,
\\
c_3(p,\omega_n,\mu,\alpha) &=  0.
\end{align*}

For small $\delta$, we have $h_i(p,\omega_n,\delta,\mu,\alpha) \simeq  b_i (p,\omega_n,\mu,\alpha) \delta^2 + c_i (p,\omega_n,\mu,\alpha) $, so at low temperatures
\begin{align*}
q_i &\simeq 4T \delta^2 \sum_{n} \int \frac{d^3 \bm{p}}{(2\pi)^3}
\frac{b_i(p,\omega_n,\mu,\alpha) }{{(\xi^2 + \omega^2_n})^4}
+
4T \sum_{n} \int \frac{d^3 \bm{p}}{(2\pi)^3}
\frac{c_i(p,\omega_n,\mu,\alpha) }{{(\xi^2 + \omega^2_n})^4}
\\&=
4T \mu \delta^2
\frac{\mathcal{N}_0}{4}
\sum_{n} \int_{-\Omega}^{\Omega}
\frac{u_i(\alpha) \xi^2 + v_i(\alpha) \omega_n^2}{{(\xi^2 + \omega^2_n})^4} d \xi
+
4T \frac{\mathcal{N}_0}{4}
w_i(\alpha)
\sum_n
 \int_{-\Omega}^{\Omega}
\frac{d \xi}{(\xi^2 + \omega^2_n)^2}
\\&=
\mu \delta^2 u_i(\alpha) \mathcal{N}_0
T \sum_{n}
\int
\frac{\xi^2 }{{(\xi^2 + \omega^2_n})^4} d\xi
+
\mu \delta^2 v_i(\alpha) \mathcal{N}_0
T \sum_{n}
\int
\frac{\omega_n^2}{{(\xi^2 + \omega^2_n})^4} d\xi
+
\mathcal{N}_0
w_i(\alpha)
T \sum_{n} \int
\frac{d \xi}{(\xi^2 + \omega^2_n)^2}
\\&=
\Big[
0.0013~u_i(\alpha)
+
0.0064~v_i(\alpha)
\Big]
\frac{\mu \delta^2 \mathcal{N}_0 }{T_c^4}
+
0.1066  w_i(\alpha)
\frac{\mathcal{N}_0}{T_c^2},
\end{align*}
where $\xi = p^2 - \mu$, $\mathcal{N}_0 = \mathcal{N}_{+}(\mu,0) + \mathcal{N}_{-}(\mu,0) = \sqrt{\mu}/\pi^2$, and $x = \xi/T$. This is Eq.~(\ref{eq_qi}) in the main text. In the last line, we utilize
\begin{align*}
T \sum_{n = -\infty}^{\infty}
\int_{-\Omega}^{\Omega}
\frac{\xi^2 }{{(\xi^2 + \omega^2_n})^4} d\xi
&=
\frac{1}{T_c^4}
\int_{0}^{\infty}
\frac{(15-6 x^2 ) \sinh x - (x^3+15x - 15 \sinh x ) \cosh x + 2 x^3 -15 x}{48 x^5 (1+\cosh x)^2} = \frac{0.0013}{T_c^4},
\\
T \sum_{n = -\infty}^{\infty}
\int_{-\Omega}^{\Omega}
\frac{\omega_n^2}{{(\xi^2 + \omega^2_n})^4} d\xi
&=
\frac{1}{T_c^4}
\int_{0}^{\infty}
\frac{3 \sinh x + (x^3 - 3 x+3 \sinh x ) \cosh x
-2 x^3 -3 x }{48 x^5 (1+\cosh x)^2} = \frac{0.0064}{T_c^4},
\\
T \sum_{n = -\infty}^{\infty}
\int_{-\Omega}^{\Omega}
\frac{d \xi}{(\xi^2 + \omega^2_n)^2}
&=
\frac{1}{T_c^2}
\int_{0}^{\infty}
\frac{\sinh x - x}{2 x^3 (1+ \cosh x )} = \frac{0.1066}{T_c^2},
\end{align*}
where we performed the integrals numerically.

\end{widetext}

\end{appendix}


\begin{thebibliography}{99}
\bibitem{remark} This was the result found in Ref. \cite{ho} for paired particles of spin 3/2.
\bibitem{comment1} This is also why in Ref. \cite{ho}, $q_3=0$ is found only for pairing of spin-3/2 particles.
\end{thebibliography}
\end{document}